\documentclass[amssymb,prb,twocolumn,showpacs,superscriptaddress]{revtex4}
\usepackage{graphicx}
\usepackage{epsfig}
\usepackage{longtable}
\usepackage{amsmath}
\usepackage{amssymb}

\begin{document}


\title{The Fe-Mg interplay and the effect of deposition mode\\
in (Ga,Fe)N doped with Mg}

\author{A.~Navarro-Quezada}
\email{andrea.navarro-quezada@jku.at}

\affiliation{Institut f\"ur Halbleiter-und-Festk\"orperphysik, Johannes Kepler University, Altenbergerstr. 69, A-4040 Linz, Austria}

\author{N.~Gonzalez Szwacki}
\affiliation{Institute of Theoretical Physics, Faculty of Physics, University of Warsaw, PL-00-681 Warszawa, Poland}

\author{W.~Stefanowicz}
\affiliation{Institute of Physics, Polish Academy of Sciences, al. Lotnik\'{o}w 32/46, PL-02-668 Warszawa, Poland}

\author{Tian Li}
\affiliation{Institut f\"ur Halbleiter-und-Festk\"orperphysik, Johannes Kepler University, Altenbergerstr. 69, A-4040 Linz, Austria}

\author{A.~Grois}
\affiliation{Institut f\"ur Halbleiter-und-Festk\"orperphysik, Johannes Kepler University, Altenbergerstr. 69, A-4040 Linz, Austria}

\author{T.~Devillers}
\affiliation{Institut f\"ur Halbleiter-und-Festk\"orperphysik, Johannes Kepler University, Altenbergerstr. 69, A-4040 Linz, Austria}

\author{R.~Jakie\l{}a}
\affiliation{Institute of Physics, Polish Academy of Sciences, al. Lotnik\'{o}w 32/46, PL-02-668 Warszawa, Poland}

\author{B.~Faina}
\affiliation{Institut f\"ur Halbleiter-und-Festk\"orperphysik, Johannes Kepler University, Altenbergerstr. 69, A-4040 Linz, Austria}

\author{J.~A.~Majewski}
\affiliation{Institute of Theoretical Physics, Faculty of Physics, University of Warsaw, PL-00-681 Warszawa, Poland}

\author{M.~Sawicki}
\affiliation{Institute of Physics, Polish Academy of Sciences, al. Lotnik\'{o}w 32/46, PL-02-668 Warszawa, Poland}

\author{T.~Dietl}
\affiliation{Institute of Physics, Polish Academy of Sciences, al. Lotnik\'{o}w 32/46, PL-02-668 Warszawa, Poland}
\affiliation{Institute of Theoretical Physics, Faculty of Physics, University of Warsaw, PL-00-681 Warszawa, Poland}

\author{A.~Bonanni}
\email{alberta.bonanni@jku.at}
\affiliation{Institut f\"ur Halbleiter-und-Festk\"orperphysik, Johannes Kepler University, Altenbergerstr. 69, A-4040 Linz, Austria}


\date{\today}

\begin{abstract}
The effect of Mg codoping and Mg deposition mode on the Fe distribution in (Ga,Fe)N layers grown by metalorganic vapor phase epitaxy is investigated. Both homogeneously- and digitally-Mg codoped samples are considered and contrasted to the case of (Ga,Fe)N layers obtained without any codoping by shallow impurities. The structural analysis of the layers by high-resolution transmission electron microscopy and by high-resolution- and synchrotron x-ray diffraction gives evidence of the fact that in the case of homogenous-Mg doping, Mg and Fe competitively occupy the Ga-substitutional cation sites, reducing the efficiency of Fe incorporation. Accordingly, the character of the magnetization is modified from ferromagnetic-like in the non-codoped films to paramagnetic in the case of homogeneous Mg codoping. The findings are discussed $\textit{vis-\`a-vis}$ theoretical results obtained by $\textit{ab initio}$ computations, showing only a weak effect of codoping on the pairing energy of two Fe cations in {\em bulk} GaN. However, according to these computations, codoping reverses the sign of the paring energy of Fe cations at the Ga-rich {\em surface}, substantiating the view that the Fe aggregation occurs at the growth surface. In contrast to the homogenous deposition mode, the {\em digital} one is found to remarkably promote the aggregation of the magnetic ions. The Fe-rich nanocrystals formed in this way are distributed non-uniformly, giving reason for the observed deviation from a standard superparamagnetic behavior.\\
\end{abstract}

\pacs{68.55.Nq, 75.75.Cd, 81.15.Gh,75.50 Pp, 61.10 Nz}
\keywords{MOVPE, Ferromagnetic, Iron nitrides, GaN}

\maketitle

\section{Introduction}
One of the principal goals of nowadays materials science is to develop tools allowing for the bottom-up assembling of nanostructures with predefined properties and functionalities.
It was theoretically suggested\cite{Dietl:2006_NM} that the codoping of magnetic semiconductors with shallow impurities affects the self-assembly of magnetic nanocrystals during epitaxy, and therefore modifies both the global and local magnetic behavior of the material. This concept was qualitatively corroborated by experimental data for (Zn,Cr)Te (Ref.\,\onlinecite{Kuroda:2007_NM}) and (Ga,Fe)N (Refs.\,\onlinecite{Bonanni:2008_PRL, Rovezzi:2009_PRB}). In the latter case it was proven that by doping (Ga,Fe)N with donors (Si) the charge state of the Fe ions is reduced from Fe$^{3+}$ to Fe$^{2+}$, affecting the binding energy of the Fe pairs, and suppressing the ions' aggregation even at a high Fe precursor flow rate.\cite{Bonanni:2008_PRL} The quenching of Fe aggregation, and therefore of either crystallographic or chemical phase-separation, results in an overall reduced magnetization of the layers, particularly at high temperatures, where the paramagnetic response of dilute spins is small.

While the effect of doping of (Ga,Fe)N with donors may be considered experimentally clarified, the role of intentional doping with acceptors (Mg) rises several remarkable issues, namely: (i) does the introduction of acceptors in a feasible concentration modify the Fe charge state to Fe$^{4+}$ or do the holes remain bound to the Mg ions?\cite{Malguth:2008_pssb,Pacuski:2008_PRL} (ii)  in either case, does the introduction of acceptors affect, similarly to donors,\cite{Bonanni:2008_PRL, Rovezzi:2009_PRB} the aggregation of Fe? (iii) which is the effect of Mg-codoping on the magnetization of the layers? (iv) is the process of aggregation essentially a bulk effect [as in annealed (Ga,Mn)As] (Refs.\,\onlinecite{De_Boeck:1996_APL}, \onlinecite{Kovacs:2011_JAP}) or it occurs at the growth surface, as suggested by our preliminary experimental\cite{Bonanni:2008_PRL} and theoretical\cite{Gonzalez:2011_PRB} results for (Ga,Fe)N? (v) can the chemistry and spatial distribution of Fe-rich nanocrystals be controlled by codoping protocols?

In the present work, we report on the effect of Mg doping and deposition mode, considering three different series of (Ga,Fe)N samples: (i) without any codoping; (ii) with Mg continuously provided during the growth process (homogeneous codoping) and (iii) with digital Mg-codoping ($\delta$Mg).\cite{Simbrunner:2007_APL}  Our x-ray diffraction (XRD), transmission electron microscopy (TEM), and superconducting quantum interference device (SQUID) data show that {\em homogeneous} Mg codoping reduces the Fe incorporation with the consequent suppression of Fe-rich nanocrystals formation, as preliminary reported.\cite{Bonanni:2008_PRL}  Interestingly, utterly different effects  are found in the case of {\em digital} codoping.  In this case, the self-assembling of Fe-rich nanocrystals is promoted, and their chemical composition and spatial distribution affected. Thus, the present analysis complements, by providing fundamental information on the role of the mode of acceptors incorporation into the system, the extensive work others\cite{Kuwabara:2001_JJAP,Ofuchi:2001_APL,Heikman:2003_JCG,Kane:2007_pssa,Malguth:2008_pssb} and we\cite{Bonanni:2008_PRL,Rovezzi:2009_PRB,Pacuski:2008_PRL,Bonanni:2007_PRB,Navarro:2010_PRB} have carried out on controlling the structural and magnetic properties of (Ga,Fe)N. The data presented here demonstrate that various modes of codoping may serve, in a bottom-up fashion, to hamper or to promote phase separation and the self-assembling of Fe-rich nanocrystals in GaN.

These experimental results are confronted with our $\textit{ab initio}$ calculations carried out according to a methodology developed and employed earlier\cite{Gonzalez:2011_PRB} to confirm the experimental observation\cite{Bonanni:2008_PRL} that the aggregation of Fe in GaN occurs at the growth surface. Other relevant first principles studies include works on the electronic states brought about by Ga-substitutional Fe ions in GaN.\cite{Rovezzi:2009_PRB,Alippi:2011_PRB} It was found that the Fe$^{3+}$ ion, in addition to acting as a mid-gap acceptor, gives rise to a hole trap level localized at 0.3\,eV above the  top of the valence band,\cite{Alippi:2011_PRB} substantiating earlier theoretical suggestions\cite{Dietl:2008_PRB} and experimental findings.\cite{Pacuski:2008_PRL}
Recently, the energetic and magnetic properties of Fe incorporated on a clean and Ga-bilayer GaN(0001) surface have been evaluated\cite{Gonzalez:2011_JAP} within a formalism similar to ours.\cite{Gonzalez:2011_PRB} It was found that a Fe atom occupies preferentially, by $\sim$2.0\,eV, a surface substitutional Ga site rather than a near-surface interstitial site. Moreover, the authors conclude that under N-rich conditions the Fe atoms incorporate into Ga substitutional sites at the top bilayer with no tendency to migrate towards the bulk.

Here, we present computed values of cohesive energies for various Fe, Mg, and also Si clusters in the bulk and at the Ga-terminated GaN(0001) surface, assuming that the Fe ions occupy Ga-substitutional positions, in agreement with the first principles studies referred to above.\cite{Gonzalez:2011_JAP} The computational data allow us to interpret the effect of codoping by either Mg or Si on the incorporation of the Fe ions. We also put forward an explanation for the highly non-random distribution of Fe-rich nanocrystals in the film volume, and relate it to the magnetic properties of this ferromagnetic/semiconductor nanocomposite system.

The manuscript is structured as follows: the experimental details on the fabrication and on the characterization methods are given in Sec.\,II, where a table listing the samples considered and their significant parameters is included. In Sec.\,III we discuss the structural and magnetic analysis of the two differently Mg-doped series of (Ga,Fe)N epilayers. We then present, in Sec.\,IV, the results of first principles computations carried out within a model developed to account for the Fe-Mg interplay observed experimentally and to compare it with the Fe-Si case. A series of often unanticipated conclusions emerging from the comparison of experimental and computational results are outlined in Sec.\,V. Within the final section, we provide a summary and an outlook prompted by the results.

\section{Samples and experimental procedure}
\subsection{Samples}

Reference (Ga,Fe)N and two series of (Ga,Fe)N layers codoped with Mg have been fabricated by metalorganic vapor phase epitaxy (MOVPE) in a 200RF AIXTRON horizontal system at a substrate temperature $T_{{\mathrm{s}}}$\,=\,850\,$^\circ$C with trimethylgallium (TMGa), ammonia (NH$_3$), ferrocene (Cp$_2$Fe) and biscyclopentadienyl magnesium (Cp$_2$Mg) as precursors. The 1\,$\mu$m thick doped layers are deposited onto a 1\,$\mu$m thick GaN buffer grown on \textit{c}-plane sapphire (Al$_2$O$_3$) substrates following the procedure described elsewhere,\cite{Bonanni:2007_PRB} employing a TMGa precursor flow rate of 12 standard cubic centimeters per minute (sccm). For the homogeneously Mg-doped layers TMGa and both doping sources Cp$_2$Mg and Cp$_2$Fe are opened simultaneously during the layer growth. The (Ga,Fe)N:$\delta$Mg layers are produced by growing 60 periods of (Ga,Fe)N layers between Mg thin layers.\cite{Simbrunner:2007_APL} The enrichment of the sample surface with nitrogen is ensured during the entire growth procedure by keeping the NH$_3$ flow rate constant. The Cp$_2$Mg source flow rate is 350\,sccm, giving -- according to secondary ion mass spectroscopy (SIMS) measurements -- a mean Mg content of $\approx$10$^{19}$\,cm$^{-3}$ ($x_{\mathrm{Mg}}$\,=\,$0.02\%$) in the $\delta$-doped samples and a concentration ten times higher in the homogenously doped ones. The Fe concentration is varied by changing the Cp$_2$Fe source flow rate from 50 to 450\,sccm. A summary of the main parameters employed to fabricate the samples considered in this work is presented in Table\,\ref{Tab:table1}.
\begin{table} [h]
\begin{ruledtabular}
\caption{\label{Tab:table1} (Ga,Fe)N references and Mg-doped (Ga,Fe)N samples with the corresponding Fe and Mg mean concentrations as evaluated $via$ SIMS. The doping method is indicated with \textsl{h} for Mg-homogeneous, and $\delta$ for Mg-digital doping.}
\label{Tab:table1}
\begin{tabular}{|c|ccccc|}
sample & type & Cp$_2$Fe & Cp$_2$Mg & $x_{\mathrm{Fe}}$ & $x_{\mathrm{Mg}}$ \\
 &   &  [sccm] & [sccm] & [\%]   & [\%] \\
\hline
S907 & ref. & 100 & 0 & $0.06$ & 0 \\
S911 & ref. & 200 & 0 & $0.11$ & 0 \\
S910 & ref. & 300 & 0 & $0.21$ & 0 \\
S994 & $\delta$ & 100 & 350 & $0.15$ & $0.02$\\
S995 & $\delta$ & 200 & 350 & $0.20$ & $0.016$\\
S993 & $\delta$ & 300 & 350 & $0.38$ & $0.016$\\
S1193 & $\delta$ & 50 & 350 & $0.08$ & $0.03$\\
S1192 & ref. & 50 & 0 & $0.08$ & 0   \\
S1406 & ref. & 300 & 0 & $0.20$ & $0$   \\
S1407 & \textsl{h} & 0 & 350 & $0.00$ & $0.15$   \\
S1402 & \textsl{h} & 50 & 350 & $0.03$ & $0.23$   \\
S1405 & \textsl{h} & 100 & 350 & $0.04$ & $0.22$   \\
S1404 & \textsl{h} & 200 & 350 & $0.07$ & $0.20$ \\
S1403 & \textsl{h} & 300 & 350 & $0.11$ & $0.19$ \\
S1415 & \textsl{h} & 400 & 350 & $0.12$ & $0.17$ \\
S1416 & \textsl{h} & 450 & 350 & $0.15$ & $0.17$ \\
S1427 & \textsl{h} & 300 & 250 & $0.12$ & $0.07$  \\
S1428 & \textsl{h} & 300 & 150 & $0.10$ & $0.01$  \\
\end{tabular}
\end{ruledtabular}
\end{table} 

\subsection{Secondary-ion mass spectroscopy}
The total Fe and Mg content in the layers has been determined $via$ secondary-ion mass spectroscopy (SIMS). An undoped GaN sample implanted with 10$^{16}$ cm$^{-2}$ of $^{56}$Fe at 270\,keV is used as reference for calibration of the system.\cite{Bonanni:2007_PRB} In the (Ga,Fe)N reference layers the total Fe concentration in the films is found to increase with the nominal Cp$_2$Fe flow rate.

\subsection{X-ray diffraction}
Structural analysis of the samples is performed by high-resolution x-ray diffraction (HRXRD), and synchrotron x-ray diffraction (SXRD). Conventional x-ray diffraction measurements are carried out on a X'Pert Pro material research diffractometer at an energy of 8\,keV with the following configuration: a hybrid monochromator and a 0.25$^\circ$ divergence slit are placed in front of the incoming beam and a PixCel detector with 1\,mm active length is used, in order to reduce the diffuse scattering around the substrate peak and to improve the resolution. Due to the high intensity beam and to the enhanced resolution obtained with the mentioned configuration the diffraction peaks originating from embedded nanocrystals are resolvable. However, to obtain an enhanced signal to noise ratio, SXRD measurements have been performed at the BM20 Rossendorf Beamline (ROBL) of the European Synchrotron Research Facility in Grenoble, France. The samples are measured in coplanar geometry using a photon energy of 10.005\,keV. Diffraction scans are acquired along the growth direction [001], and in a range between the (002) diffractions of GaN and the (006) of Al$_2$O$_3$.\cite{Navarro:2010_PRB}

\subsection{Transmission electron microscopy}

The high-resolution transmission electron microscopy (HRTEM) studies are performed on cross-sectional samples prepared by standard mechanical polishing followed by Ar$^{+}$ ion milling at 4\,kV for about 1\,h. Conventional diffraction contrast images in bright-field imaging mode and high-resolution phase contrast pictures have been obtained from a JEOL 2011 Fast TEM microscope operating at 200\,kV and capable of an ultimate point-to-point resolution of 0.19\,nm and allowing to image lattice fringes with a 0.14\,nm resolution. Additionally, energy dispersive spectroscopy (EDS) analysis \textit{via} an Oxford Inca EDS equipped with a silicon detector provides information on the local composition. Selected area electron diffraction (SAED) and fast Fourier transform (FFT) procedures are employed to study the scattering orders and the $d$-spacing for respectively the larger and the smaller embedded nanocrystals.

\subsection{SQUID magnetometry}
\label{SQUID}

The magnetic moment of the samples has been measured in a Quantum Design MPMS XL 5 SQUID magnetometer between 1.85 and 400\,K and up to 5\,T following the methodology  described elsewhere.\cite{Stefanowicz:2010_PRB,Sawicki:2011_SST}

As demonstrated previously,\cite{Bonanni:2008_PRL,Pacuski:2008_PRL,Navarro:2010_PRB,Bonanni:2007_PRB} the magnetic response of (Ga,Fe)N layers can be considered as representative for semiconductors containing transition metal ions at a concentration near the solubility limit. In particular, it contains a sizable diamagnetic component from the sapphire substrate, which is carefully subtracted according to the procedure detailed recently.\cite{Stefanowicz:2010_PRB,Sawicki:2011_SST}
The remaining magnetic moment $m(H)$ is: (i) purely paramagnetic and assigned to substitutional Fe$^{3+}$ ions if the system is dilute, while (ii) it is build up from the paramagnetic part as well as from a ferromagnetic-like component and from an antiferromagnetic contribution linear in $H$ -- these latter, both stemming from various magnetically ordered nanocrystals with high Fe content\cite{Navarro:2010_PRB} -- if the material undergoes phase-separation.

Due to the fact that the Fe-rich nanocrystals are characterized by relatively high ordering temperatures, a significant variation of $m(H)$ at low temperatures allows us to assess quite precisely the concentration of paramagnetic ions $x_{\text{para}}$. Following the procedure employed previously,\cite{Pacuski:2008_PRL,Navarro:2010_PRB} we obtain $x_{\text{para}}$ by fitting $g\mu_{\text{B}}Sx_{\text{para}}N_0 \Delta$B$_S(H,T)$ to the difference between the experimental values of the magnetization measured at 1.85 and 5.0 K, where $\Delta$B$_S(H,T)$ is the difference of the corresponding paramagnetic Brillouin functions $\Delta$B$_S(H,T)$\,=\,B$_S(H, 1.85\,\mathrm{K})$-B$_S(H, 5\,\mathrm{K})$. We adopt here parameters suitable for Fe$^{3+}$ ions in GaN, {\em i.e.}, the spin $S$=\,$5/2$, the corresponding Land\'e factor $g$\,=\,$2.0$, and the cation density $N_0$\,=\,4.4\,$\times$\,10$^{22}$\,cm$^{-3}$, treating  $x_{\text{para}}$ as the only fitting parameter. Its value is then employed to calculate the paramagnetic contribution at any other temperature according to $M(T,H)\,=\,g\mu_{\text{B}}Sx_{\text{para}}N_0$B$_S(H,T)$, which is then subtracted from the experimental data to obtain the magnitude of the magnetization $M_{\text{ferro}}(H,T)$ and its saturation value from which we establish $x_{\text{ferro}}$, a lower limit of the Fe concentration in the Fe-rich nanocrystals. We assume the magnetic moment per Fe ion to be $Sg\mu_{\text{B}}$\,=\,2.0\,$\mu_{\text{B}}$, the value determined for $\epsilon$-Fe$_3$N.\cite{Navarro:2010_PRB}  Since, in general, the distribution of ferromagnetic nanocrystals is highly non-uniform, $x_{\text{ferro}}$ is intended as a mean value.

\subsection{Electron spin resonance}
\label{ESR}

Electron spin resonance (ESR) spectra have been obtained with a Bruker Elexsys E 500 spectrometer operating in the microwave X-band ($\approx 9.8$\,GHz) equipped with a pumped helium cryostat.  The measurements have been performed at a temperature of 2\,K at the microwave power of $\approx 2$\,mW.  The magnetic field is applied parallel to the crystal $c$-axis. To allow for lock-in detection, its magnitude is modulated at 100\,kHz with an amplitude of 5\,Oe.  The integrated amplitude of the Fe$^{3+}$ ESR signal, omitting the region in the immediate vicinity of residual resonances corresponding to $g$\,=\,$2.0$, serves as a relative measure of the concentration of Fe$^{3+}$ ions.\cite{Bonanni:2007_PRB}

\section{Experimental results}

We discuss here the results of the SIMS, XRD, TEM, SQUID, and ESR measurements carried out according to the methodologies specified in the previous section. The findings provide information on the Fe incorporation, distribution, concentration, and on the corresponding magnetic properties of the considered samples. In order to demonstrate that we can control the Fe incorporation and distribution by selecting an appropriate codoping mode, the experimental results for three series of samples, namely: (Ga,Fe)N (reference samples), (Ga,Fe)N:Mg (homogeneous Mg-codoping), and (Ga,Fe)N:$\delta$Mg (digital Mg-codoping) are compared.

\subsection{Fe incorporation -- SIMS data}

\begin{figure}[ht]
	\includegraphics[width=\columnwidth]{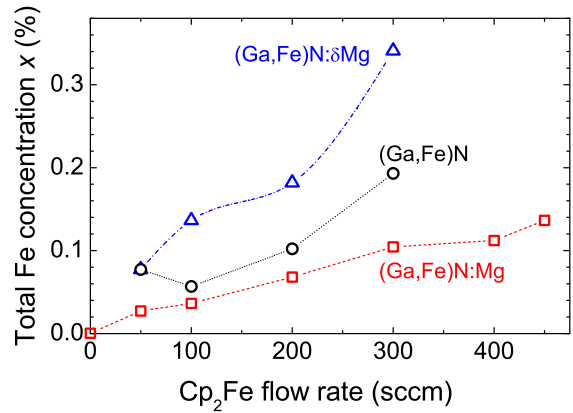}
	\caption{Total Fe content as evaluated from SIMS measurements as a function of the Fe-source flow-rate for the reference (Ga,Fe)N and for the correspondent (Ga,Fe)N:Mg $x_{\mathrm{Mg}}$\,=\,0.2\% layers, as determined from SIMS.}
	\label{fig:SIMS_Fe}
\end{figure}

\begin{figure}[ht]
		\centering
        \includegraphics[width=\columnwidth]{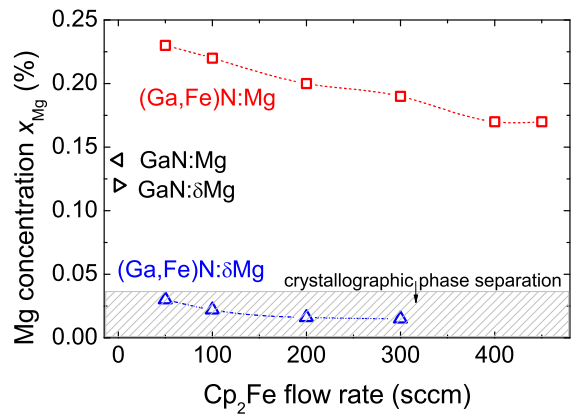}
      \caption{Mg concentration as determined by SIMS $vs.$ Cp$_2$Fe source flow for (Ga,Fe)N:Mg (squares) and (Ga,Fe)N:$\delta$Mg (triangles). GaN:Mg reference samples [homogeneous ($\lhd$) and digital ($\rhd$)] are included. All layers with $x_{\mathrm{Mg}} \leq 0.03\%$ show Fe crystallographic phase separation for Cp$_2$Fe source flows as low as 50 sccm.}
    \label{fig:SIMS_Mg}
\end{figure}

In Fig.\,\ref{fig:SIMS_Fe} the Fe concentration as obtained from SIMS data averaged over the layer thickness is reported.  As seen, the two employed Mg codoping modes have a considerable, but reverse effect on the Fe incorporation, that is on the total Fe concentration in the layers for a given Fe precursor flow rate. In particular, homogeneous Mg-codoping reduces substantially the Fe concentration, which is -- on the other hand -- significantly enhanced in case of digital Mg-codoping in comparison to the (Ga,Fe)N reference films. 
It is interesting to establish how the effects described above evolve with the Mg concentration. According to the SIMS data presented in Fig.\,\ref{fig:SIMS_Mg} only above a certain threshold of the Mg concentration, namely for $x_{\mathrm{Mg}} \geq 0.03\%$, Mg appears to prevail in thwarting the efficient incorporation of the magnetic ions.

\subsection {Fe-rich nanocrystals -- XRD and TEM data}

\begin{figure}[ht]
	\includegraphics[width=\columnwidth]{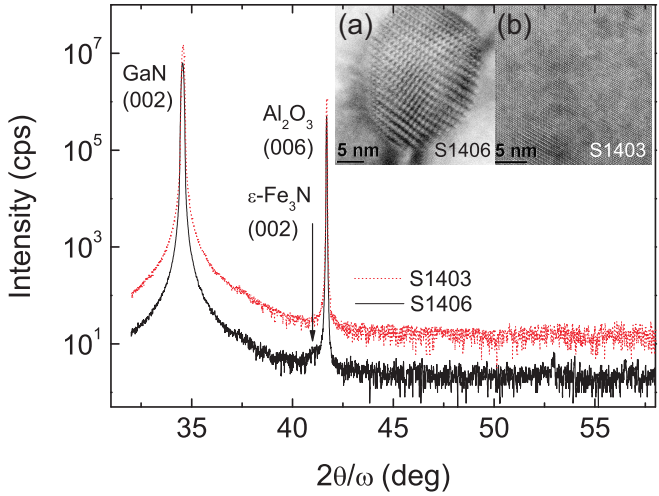}
	\caption{HRXRD curves for a reference (Ga,Fe)N grown at Fe precursor flow rate of 300\,sccm with $\varepsilon$-Fe$_3$N embedded nanocrystals and for the corresponding dilute (Ga,Fe)N:Mg layer with $x_{\mathrm{Mg}}$\,=\,0.2\%. Insets: HRTEM images - (a): $\varepsilon$-Fe$_3$N embedded in the (Ga,Fe)N reference sample; (b) no precipitates in the corresponding (Ga,Fe)N:Mg layer.}
	\label{fig:MRD_clean_homo}
\end{figure}

As it can be seen in Fig.\,\ref{fig:MRD_clean_homo}, in (Ga,Fe)N films, grown at a sufficiently high Fe precursor flow rate of 300\,sccm, there are signatures of crystallographic phase separation witnessed by the HRXRD peaks corresponding to the expected $\varepsilon$-Fe$_3$N secondary phase, detected also by HRTEM, as  displayed in the inset to Fig.\,\ref{fig:MRD_clean_homo}. In contrast, in the correspondent (Ga,Fe)N:Mg sample grown with the same parameter, but homogeneously doped with Mg, the Fe ions are less efficiently incorporated and the sample is dilute, as confirmed in Fig.\,\ref{fig:MRD_clean_homo} by the absence of both additional diffraction peaks in the HRXRD spectra and of nanocrystals in the HRTEM images.

\begin{figure}[ht]
	\includegraphics[width=\columnwidth]{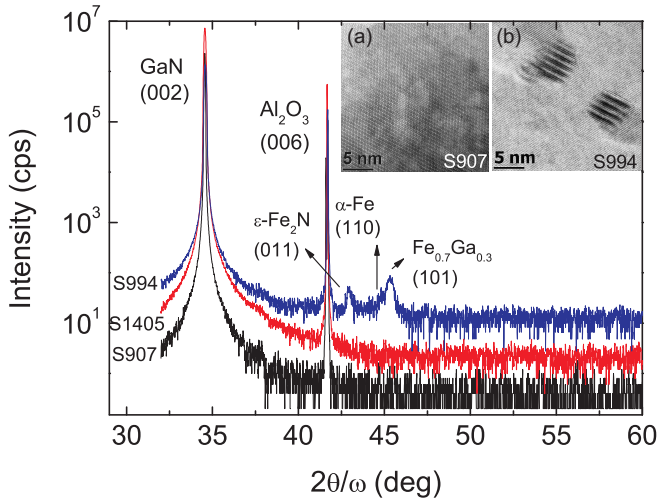}
	\caption{HRXRD spectra for a reference dilute (Ga,Fe)N sample and for corresponding layers homogeneously and digitally codoped by Mg, respectively. All films were grown at the Fe precursor flux rate of 100 sccm.}
	\label{fig:SXRD_clean_vs_delta}
\end{figure}

In Fig.\,\ref{fig:SXRD_clean_vs_delta}, the HRXRD spectrum for a reference dilute (Ga,Fe)N sample grown at the relatively low Fe precursor flow rate of 100\,sccm, is compared to those obtained for a homogeneously- and for a digitally-Mg codoped layer grown under the same Fe flow rate. While for the (Ga,Fe)N reference and for the homogeneously Mg codoped (Ga,Fe)N:Mg layer there is no evidence of crystallographic phase separation and only the diffraction peaks from the GaN buffer and from the sapphire substrate are detected, the (Ga,Fe)N:$\delta$Mg sample gives clear evidence of diffraction arising from secondary phases.

\begin{figure}[ht]
	\includegraphics[width=\columnwidth]{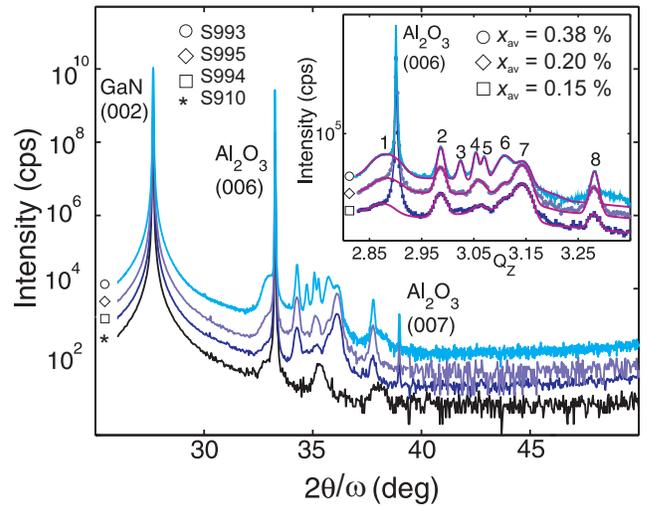}
	\caption{SXRD spectra of highly Fe-doped (Ga,Fe)N (*) and (Ga,Fe)N:$\delta$Mg samples with different Fe concentrations (0.15\%$\leq$x$\leq$0.38\%) giving evidence of the presence of secondary Fe-rich phases in the layers. Inset: detail of diffraction peaks from the secondary phases -- gaussian fit of the peaks. The phases corresponding to the diffraction peaks labelled 1 to 8 are listed in Table\,\ref{tab:param}.}
	\label{fig:SXRD}
	\end{figure}

In order to identify the detected Fe-rich phases, SXRD measurements have been carried out on a series of (Ga,Fe)N:$\delta$Mg samples fabricated with constant nominal $\delta$Mg-doping and by varying the nominal Fe concentration throughout the series. In Fig.\,\ref{fig:SXRD} the SXRD spectra for these samples are reported together with the corresponding (Ga,Fe)N phase-separated (nominal Fe content above 0.2\%) reference.  Several diffraction peaks close to the Al$_2$O$_3$ (006) peak are observed and are labeled from 1 to 8 in the inset to Fig.\,\ref{fig:SXRD}. The Fe-rich phases in the (Ga,Fe)N phase-separated sample are all observed in the Mg-codoped ones, however their density is higher in the $\delta$Mg-doped layers as inferred from the higher peak intensity and the narrower full-width-at-half-maximum (FWHM) of the peaks from the nanocrystals in the $\delta$Mg-doping case compared with those in (Ga,Fe)N. The identified Fe-rich phases, their \textit{d}-spacing and their size, as obtained directly from the FWHM of the SXRD peaks,\cite{Navarro:2010_PRB} are summarized in Table\,\ref{tab:param}. Some of the phases identified, like $\textit{e.g.}$ the $\textit{galfenol}$ Fe$_{1-x}$Ga$_{x}$ ones, are not present in phase-separated (Ga,Fe)N when not $\delta$Mg-doped.

\begin{table}[ht]
\caption{Diffraction peaks from Fe-rich phases in (Ga,Fe)N:$\delta$Mg, as observed in Fig.\,\ref{fig:SXRD}}
\label{tab:param}
\begin{tabular}{|c|cccc|}
\hline
\hline
peak  & \textit{d}-spacing & av. size & phase & (hkl) \\
 Nr.    & (nm)       & (nm) &  &  \\
\hline
1 & 0.2188 & 13 & $\varepsilon$-Fe$_3$N & (002) \\
2 & 0.2102 & 43 & $\varepsilon$-Fe$_2$N & (011) \\
3 & 0.2077 & 60 & $\varepsilon$-Fe$_3$N & (111) \\
4 & 0.2057 & 64 & Fe$_3$Ga & (220) \\
5 & 0.2048 & 38 & $\varepsilon$-Fe$_{2.4}$N & (221) \\
6 & 0.2020 & 25 & $\alpha$-Fe & (110) \\
7 & 0.1998 & 25 & Fe$_{0.7}$Ga$_{0.3}$ & (101) \\
8 & 0.1914 & 24 & $\gamma$'-Fe$_4$N & (200) \\
\hline
\hline
\end{tabular}
\end{table}

The FFT of the Moir{\'e} fringes in the HRTEM images of selected nanocrystals, reveals the presence of mainly $\varepsilon$-Fe$_3$N in the (Ga,Fe)N reference and a variety of Fe-rich phases with \textit{d}-spacings as reported from SXRD in the (Ga,Fe)N:$\delta$Mg ones. The EDS spectra show that the Fe signal is significantly enhanced in the region of the nanocrystals, confirming their Fe-rich nature.

\begin{figure}[ht]
	\includegraphics[width=0.9\columnwidth]{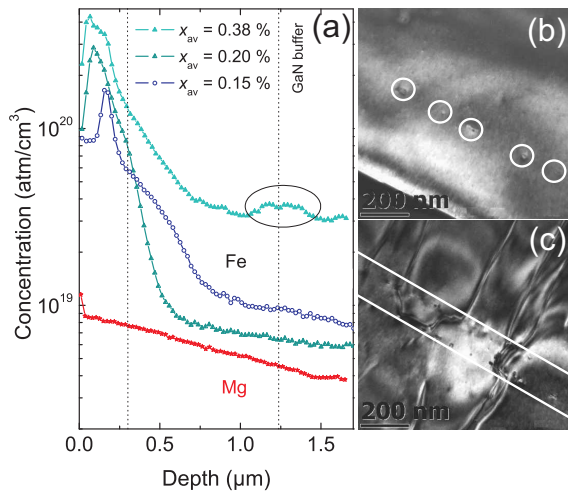}
	\caption{(a) SIMS profiles of (Ga,Fe)N:$\delta$Mg proving the inhomogeneous distribution of Fe in the layers. (b),(c): TEM bright-field images of nanocrystals in the GaN:Fe:$\delta$Mg sample S993 ($x_{\mathrm{Fe}}$\,=\, 0.38\%) -- (b) close to the sample surface; (c) at the interface GaN-buffer/(Ga,Fe)N:$\delta$Mg-layer.}
	\label{fig:tem}
\end{figure}

Another important issue concerns the spatial distribution of the Fe-rich nanocrystals. SIMS measurements on the $\delta$Mg-doped samples evidence an inhomogeneous distribution of Fe in the layers along the growth direction, as seen in Fig.\,\ref{fig:tem}(a). Up to 300\,nm below the surface the Fe concentration reaches its maximum for all $\delta$Mg-doped samples. Additionally, in the sample containing 0.38\% of Fe, an increased concentration of the magnetic ions is resolved close to the interface. This is also observed in TEM images, where in the highly $\delta$Mg-doped layer the nanocrystals are found 300\,nm below the sample surface -- as seen in Fig.\,\ref{fig:tem}(b) -- in two narrow regions separated by 180\,nm and close to the  interface -- as evidenced in Fig.\,\ref{fig:tem}(c). In the undoped (Ga,Fe)N sample, the nanocrystals lie around 200\,nm below the sample surface.\cite{Li:2008_JCG2, Navarro:2010_PRB}

\subsection{Magnetic properties and Fe concentrations -- SQUID data}

A strict correlation between the presence of Fe-rich nanocrystals and the emergence of high temperature ferromagnetism has already been demonstrated for (Ga,Fe)N (Ref.\,\onlinecite{Bonanni:2008_PRL,Bonanni:2007_PRB,Navarro:2010_PRB}). In fact, the SQUID data can serve to quantify the concentrations of Fe ions contributing to ferromagnetic and paramagnetic signals, respectively, as described in Sec.\,\ref{SQUID}.

Since the concentrations of Fe and Mg are similar, the question arises about a contribution of the Mg impurities to the magnetic response. From ESR measurements, we can state that no paramagnetic signal originates from holes bound to Mg acceptors in GaN:Mg, presumably because of compensation by hydrogen centers. Likely due to the same reason, the paramagnetic signal is -- within 3$\%$ -- independent of the orientation of the $c$-axis of the samples with respect to the magnetic field, indicating that the Fe ions retain the 3+ charge state also in (Ga,Fe)N:Mg. 
\begin{figure}[ht]
	\includegraphics[width=\columnwidth]{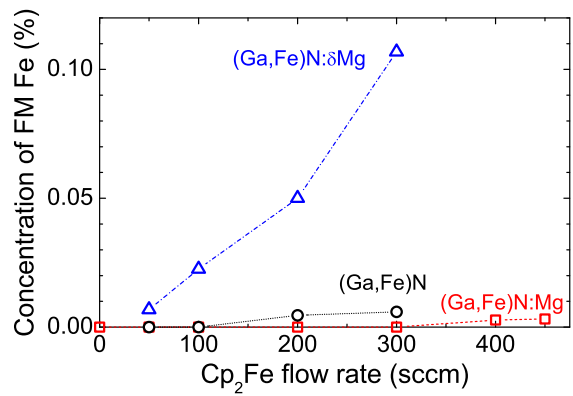}
	\caption{(Color online) Mean concentrations of Fe ions contributing to ferromagnetic response as a function of Fe flux rate, as measured by SQUID for (Ga,Fe)N (circles); (Ga,Fe)N:Mg (squares); (Ga,Fe)N:$\delta$Mg (triangles). Dashed lines are guides for the eye.}
	\label{fig:ferro}
\end{figure}

\begin{figure}[ht]
	\centering
	\includegraphics[width=\columnwidth]{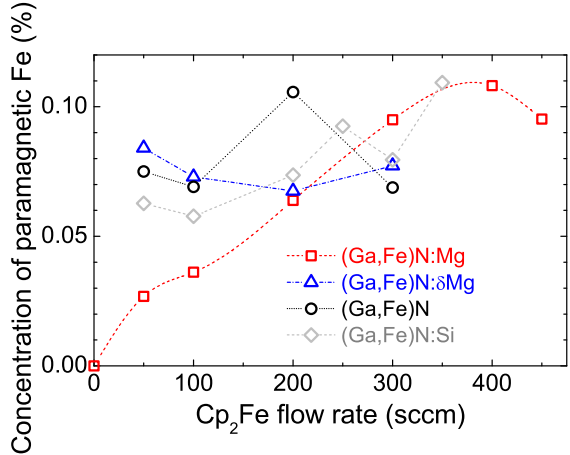}
	\caption{(Color online) Mean concentrations of Fe ions contributing to the paramagnetic response as a function of the Fe flow rate, as measured by SQUID for (Ga,Fe)N (circles); (Ga,Fe)N:Mg (squares); (Ga,Fe)N:$\delta$Mg (triangles). Results obtained earlier for (Ga,Fe)N:Si (Ref.\,\onlinecite{Bonanni:2008_PRL}) are shown for comparison (diamonds). Dashed lines are guides for the eye.}
	\label{fig:para}
\end{figure}

In Figs.\,\ref{fig:ferro} and \ref{fig:para}
the amount of Fe contributing to the ferromagnetic and paramagnetic signals, is respectively given. We see that the onset of ferromagnetism, which occurs at about 100\,sccm of the Fe-precursor-flow-rate in the case of (Ga,Fe)N, is shifted to up to 300\,sccm and down to 50\,sccm for homogeneous and digital Mg codoping, respectively. Thus, the present data supports the assignment of ferromagnetism to the presence of Fe-rich nanocrystals revealed by XRD and TEM, as discussed in the previous subsection. Moreover, they demonstrate directly how a particular mode of codoping can serve to control high temperature ferromagnetism in (Ga,Fe)N. 

As seen by comparing Figs.\,\ref{fig:ferro} and \ref{fig:para}, in the region where ferromagnetism sets in, the concentration of paramagnetic Fe ions saturates and then decreases. It is also clear that homogeneous Mg-codoping reduces not only the ferromagnetic contribution but also the magnitude of the paramagnetic signal, in accordance with the induced quenching of the efficiency of Fe-incorporation. Interestingly, the data obtained earlier for (Ga,Fe)N:Si (Ref.\,\onlinecite{Bonanni:2008_PRL}), and displayed also in Fig.\,\ref{fig:para}, show that Si codoping, in contrast to Mg codoping , enhances the concentration of paramagnetic Fe ions, at least at low Fe flow rates.  

\begin{figure}[ht]
	\centering
	\includegraphics[width=\columnwidth]{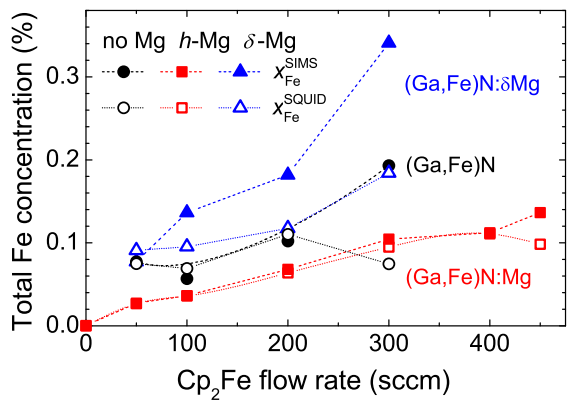}
	\caption{(Color online) Total concentrations of Fe ions determined by SIMS (full symbols) and obtained as a sum of the ferromagnetic and paramagnetic contributions determined by SQUID magnetometry (open symbols) for (Ga,Fe)N (circles); (Ga,Fe)N:Mg (squares); (Ga,Fe)N:$\delta$Mg (triangles). Dashed and dotted lines are guides for the eye.}
	\label{fig:SIMS_SQUID}
\end{figure}

In Fig.\,\ref{fig:SIMS_SQUID}, the total Fe concentration from SIMS is compared to the sum of ferromagnetic and paramagnetic components. It is seen that the concentration from SIMS is systematically higher in the ferromagnetic region, indicating that some Fe ions are not captured by the employed analysis of the SQUID data in this regime. We assign the missing contribution to the presence of antiferromagnetic Fe-rich nanocrystals stabilized by the host.\cite{Navarro:2010_PRB}

\begin{figure}[ht]
		\centering
        \includegraphics[width=\columnwidth]{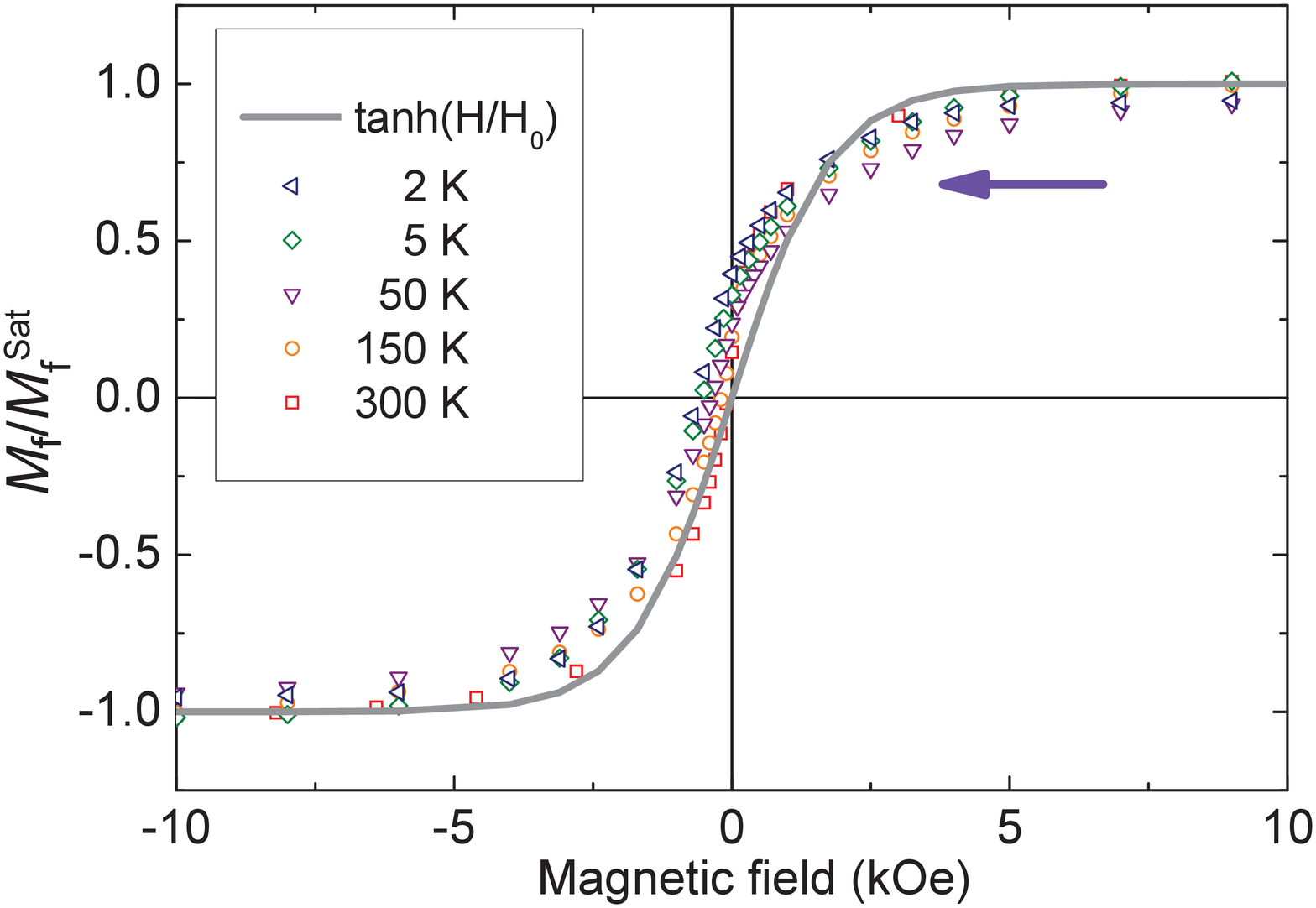}
      \caption{(Color online) Normalized magnetization at various temperatures of sample S995 [(Ga,Fe)N:$\delta$Mg, $x_{\mathrm{Fe}}$\,=\,0.2\%] with embedded nanocrystals. Points -- experimental data; solid line –- fit a Langevin-type function with $H_0$\,=\,800\,Oe. Arrow: sweep direction of the magnetic field.}
    \label{fig:magnetization_NC}
\end{figure}

The clear correlation between the appearance of ferromagnetic-like features and the presence of Fe-rich nanocrystals referred to above substantiates the fact that Fe aggregation accounts for the high temperature ferromagnetism of (Ga,Fe)N. As shown in Fig.\,\ref{fig:magnetization_NC}, the ferromagnetic component of the magnetization $M_{\text{f}}(H)$ of (Ga,Fe)N exhibits the features characteristic for high temperature ferromagnetic semiconductors. In particular, for any orientation of the magnetic field, the saturation magnetization is much larger than the spontaneous one. However, the spontaneous magnetization persists to above room temperature, indicating that for some nanocrystals the blocking temperature is higher than 300\,K. Actually, the shape of the normalized $M(H)$, particularly the presence of a small but non-zero spontaneous magnetization even at 300\,K, indicates that for some nanocrystals the superparamagnetic limit has not been reached, whereas for others $M_{\text{f}}$ is described by a temperature independent Langevin-type function:\cite{Coey:2010_NJP}
\begin{equation}
M_{\text{f}}/M^{\text{Sat}}_{\text{f}} = \tanh(H/H_{0}),
\label{eq:Coey}
\end{equation}
where $H_{0}$ is a sample dependent parameter, evaluated here to be 800\,$\pm$\,200\,Oe.

It has been suggested\cite{Coey:2010_NJP} that dipole-dipole interactions between densely packed magnetic constituents may lead to a magnetization that can be accurately parameterized by a temperature independent Langevin-type function (Eq.\,\ref{eq:Coey}),  with 4$\pi H_0 \gg M^{\text{Sat}}_{\text{f}}$ constituting the magnitude of the local magnetization.  In particular, these interactions may result in a domain or vortex structure within individual magnetic nanocrystals, that reduces their magnetization in weak magnetic fields. Furthermore, as evidenced by the TEM and SIMS measurements discussed in the previous subsection and illustrated in Fig.\,\ref{fig:tem} the distribution of magnetic nanocrystals is highly non-uniform, promoting their dipole-dipole interactions.

\subsection{Concentration of Fe$^{3+}$ ions -- ESR data}

In order to asses the microscopic origin of the paramagnetic component we have carried out ESR measurements. As in the previous studies\cite{Bonanni:2007_PRB} we observe a spectrum that is assigned to non-interacting Fe$^{3+}$ ions in a wurtzite crystal field. As shown in Fig.\,\ref{fig:ESR_deltaMg}, the Fe concentration determined from the ESR signal intensity scales consistently in both (Ga,Fe)N and (Ga,Fe)N:$\delta$Mg with the concentration of paramagnetic ions drawn from the SQUID data.  At the same time, however, the ESR signal is dramatically reduced in (Ga,Fe)N:Mg.  

The integrated peak amplitude of the $\mathrm{Fe}^{3+}$ magnetic transitions at the magnetic field parallel to the crystal $c$-axis serves as a relative measure for the concentration of these isolated paramagnetic ions,\cite{Bonanni:2007_PRB} calculated according to the following procedure: since the spectra are collected by means of a lock-in technique, they are proportional to the first derivative of the absorption with respect to the magnetic field.  

First a linear background is subtracted from the spectra which is afterwards numerically integrated to yield a signal proportional to the absorption.  These integrated spectra usually show a broad background, which is removed by substracting a polynomial of an order between 4 and 9.  Then the peaks of interest are integrated again to get their integrated intensity, the integration ranges being chosen to be the same for all samples.  These integrated peak areas are proportional to the total number of spins and are therefore normalized by the sample volume.   

To convert these integrated peak areas into absolute concentration values, they are multiplied by a calibration factor (the same for all samples), which is chosen such that it minimizes the sum over all samples of the squares of the differences between the $\mathrm{Fe}^{3+}$ concentration values obtained by ESR and SQUID according to the procedure described in Sec.\,\ref{SQUID}. The formula that is minimized is $E=\sum\limits_i \left(a I_{{\mathrm{ESR}}_i}-x_{{\mathrm{SQUID}}_i}\right)^2$, where $a$ is the calibration factor, $I_i$ is the integrated peak intensity for sample $i$ obtained from ESR and $x_i$ is the $\mathrm{Fe}^{3+}$ concentration in sample $i$ as obtained by SQUID.  The minimization leads to $a=\left(\sum\limits_i I_{{\mathrm{ESR}}_i} x_{{\mathrm{SQUID}}_i}\right)/\left(\sum\limits_i I_{{\mathrm{ESR}}_i}^2\right)$.

The Fe transition at $g \approx 2$ has not been employed for quantification, since it overlaps with a signal stemming from impurities in the substrate and varying in intensity from wafer to wafer.

\subsubsection{(Ga,Fe)N:Mg}

\begin{figure}[ht]
		\centering
        \includegraphics[width=\columnwidth]{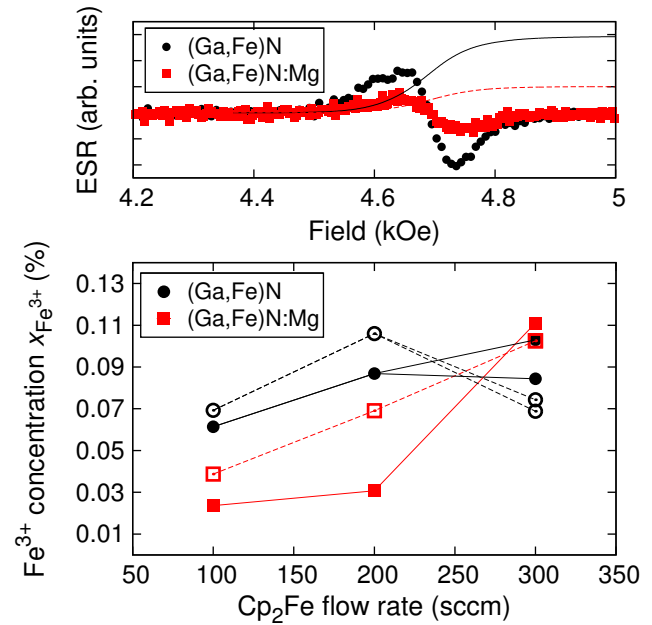}
      \caption{(Color online) ESR for (Ga,Fe)N:Mg (squares) and reference (Ga,Fe)N (circles). Upper panel -- one of the selected $\mathrm{Fe}^{3+}$ transition and double integral for the samples grown with 200 sccm $\mathrm{Cp_2Fe}$ flow rate: (Ga,Fe)N (circles, solid line) and a (Ga,Fe)N:Mg (squares, dashed line), respectively.  Lower panel -- $\mathrm{Fe}^{3+}$ concentration, as determined by ESR (filled symbols) and SQUID (open symbols).}
    \label{fig:ESR_Mg}
\end{figure}

As seen in Fig.\,\ref{fig:ESR_Mg}, the concentration of paramagnetic isolated $\mathrm{Fe}^{3+}$ as determined by ESR is consistent with the values obtained by fitting the SQUID magnetisation curves according to Sec.\,\ref{SQUID}. This agreement corraborates our statement, that most of the present Fe is indeed isolated and in the 3+ valence state. By comparing the ESR spectra of (Ga,Fe)N with those of (Ga,Fe)N:Mg, it is observed that the concentration of $\mathrm{Fe}^{3+}$ is generally decreased when homogeneously codoping (Ga,Fe)N with Mg.  

\subsubsection{(Ga,Fe)N:$\delta$Mg}

\begin{figure}[ht]
		\centering
        \includegraphics[width=\columnwidth]{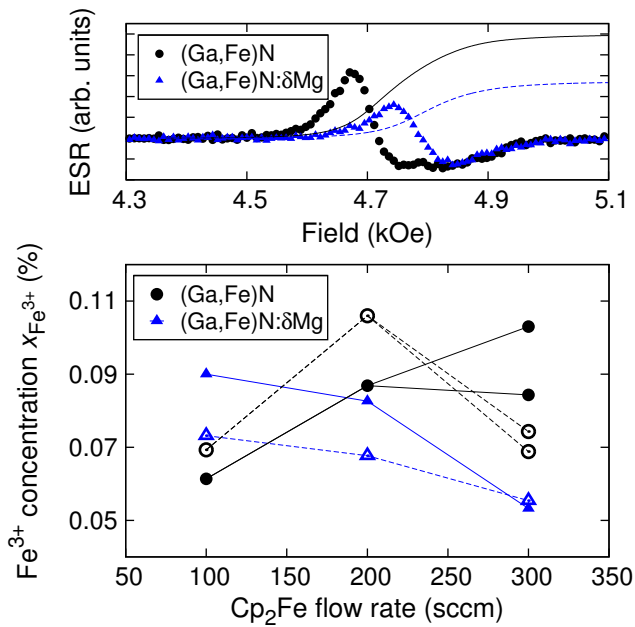}
      \caption{(Color online) ESR for (Ga,Fe)N:$\delta$Mg (triangles) and reference (Ga,Fe)N (circles). Upper panel -- selected $\mathrm{Fe}^{3+}$ transition and double integral for a digitally Mg-codoped (triangles, dashed line) and for a reference (Ga,Fe)N (circles, solid line) grown with 300 sccm $\mathrm{Cp_2FE}$ flow rate. Lower panel -- $\mathrm{Fe}^{3+}$ concentration as determined by ESR (filled symbols) and SQUID (open symbols).}
    \label{fig:ESR_deltaMg}
\end{figure}

Also for the digitally Mg-doped layers a significant consistency between the SQUID and ESR values for $\mathrm{Fe}^{3+}$ is found and summarized in Fig.\,\ref{fig:ESR_deltaMg}. In the $\delta$Mg doped samples, the concentrations of $\mathrm{Fe}^{3+}$ are systematically and notably lower than the overall Fe concentration as determined by SIMS, in agreement with the observation that a substantial amount of Fe is incorporated in secondary Fe-rich phases. Furthermore, we generally observe that the $\mathrm{Fe}^{3+}$ concentration as determined by ESR is lower in the digitally doped samples than in the reference (Ga,Fe)N.

\section{Computational results}

As mentioned earlier, the computation of the cohesive energies for various Fe, Mg, and also Si clusters in the bulk and at the Ga-terminated GaN(0001) surface allows us to interpret the effect of codoping by either Mg or Si on the incorporation of the Fe ions, especially in the frame of the different codoping modes. 

\subsection{Methodology}
The computations have been performed within the density functional theory-generalized gradient approximation DFT-GGA+\textit{U} approach with the Perdew-Burke-Ernzerhof exchange and correlation functional\cite{Perdew:1996_PRL} and ultrasoft pseudopotentials\cite{Misc1:2011} to describe electron-ion interactions as implemented in the plane wave basis Quantum Espresso code.\cite{Giannozzi:2009_JPCM}  The energy cut-off is set to 30 and 180\,Ry for the kinetic energy and electronic density, respectively. The value of the \textit{U} parameter is chosen to be 4.3\,eV for the Fe $3d$ orbitals, a similar value being employed in our previous work.\cite{Gonzalez:2011_PRB}

As previously,\cite{Gonzalez:2011_PRB} we use the supercell approach to calculate the formation energy of Fe$_n$Mg$_{\mathrm{m}}$ and Fe$_n$Si$_m$ ($n, m \leq 2$) clusters in the zinc-blende (zb) GaN and at the Ga terminated (0001) wurtzite (wz) GaN surface. The zb supercell consists of 64 ($2a \times 2a \times 2a$) atoms. A $4 \times 4 \times 4$ Monkhorst-Pack \textbf{k}-point sampling mesh is employed. The modeling at the Ga surface is carried out with a supercell containing 4\,GaN bilayers (48\,Ga and 48\,N atoms) and a vacuum region of approximately 1\,nm. The first GaN bilayer is fixed in the appropriate bulk configuration. The unit cell is taken to be  $2\sqrt{3}a \times 3a$, and the dangling bonds on the opposite nitrogen surface are saturated with 12\,H atoms. The integration over the Brillouin zone (BZ) is performed using a $4 \times 4 \times 4$ Monkhorst-Pack \textbf{k}-point mesh. We have used the Methfessel-Paxton smearing for the integration over the BZ with a smearing width of 0.05 and 0.005 for the surface and bulk calculations, respectively.
The atomic positions within the supercells have been fully optimized in all cases. The lattice constants are taken from our previous work\cite{Gonzalez:2011_PRB} and set to 0.4549\,nm for zb-GaN, and $a$\,=\,0.3211\,nm and $c$\,=\,0.5231\,nm for wz-GaN.

We have studied the formation of Fe$_n$Mg$_m$ and Fe$_n$Si$_m$ ($n, m \leq 2$) complexes in the zb-GaN matrix and also at the (0001) wz-GaN surface. The atoms in the studied clusters occupy substitutional positions in the cation sublattice of GaN. The most stable configuration for the Fe-X (X\,=\,Mg or Si) cluster is the one where Mg or Si is the nearest neighbor (NN) of the Fe atom. If a second Mg or Si is at the NN distance to Fe, then the three atoms form an almost linear X-Fe-X cluster, where X\,=\,Mg or Si. A cluster with two Fe has been previously considered\cite{Gonzalez:2011_PRB} and the stable configuration has the two transition metal atoms as NNs, both in the bulk and at the surface. The Fe$_2$X (X\,=\,Mg or Si) complex forms an isosceles triangle. Interestingly, the same configuration has been recently found to be the most stable structure for the isolated Fe$_2$Si cluster.\cite{Yu:2011_AdMatRes} Finally, the Fe$_2$X$_2$ (X\,=\,Mg or Si) cluster forms two isosceles triangles with a common Fe-Fe base. The two triangles are almost coplanar and the cluster has Cs symmetry. It should be pointed out that there is no significant difference in the shape of the structures in bulk compared to those at the surface.

\subsection{Results of {\em ab initio} studies}
As previously reported,\cite{Gonzalez:2011_PRB} two Fe ions tend to form pairs both in the bulk and at the Ga terminated surface as the heat of reaction (pairing energy $E_{\text{d}}$) for those clusters is \,-0.158 and  -0.125\,eV, respectively. The formation of Fe-Mg and Fe-Si cation pairs in bulk GaN is energetically favorable since $E_{\text{d}}$ for those clusters is  -0.178 and -1.349\,eV, respectively. In contrast, at the Ga surface only Fe and Si tend to form pairs ($E_{\text{d}}$\,=\,-0.309\,eV), whereas Fe and Mg avoid being at NNs positions ($E_{\text{d}}=0.187$\,eV). It is worth mentioning that the $E_{\text{d}}$ value for Fe-Si pairs in bulk is similar to the binding energy -1.44\,eV for an isolated Fe-Si cluster.\cite{Yu:2011_AdMatRes}  Finally, neither Mg nor Si are likely to form monoatomic pairs in the bulk ($E_{\text{d}}$\,=\,0.102 and 0.847\,eV, respectively) or at the surface ($E_{\text{d}}=0.130$ and 0.067\,eV, respectively).

\begin{figure}[ht]
		\centering
        \includegraphics[width=0.8\columnwidth]{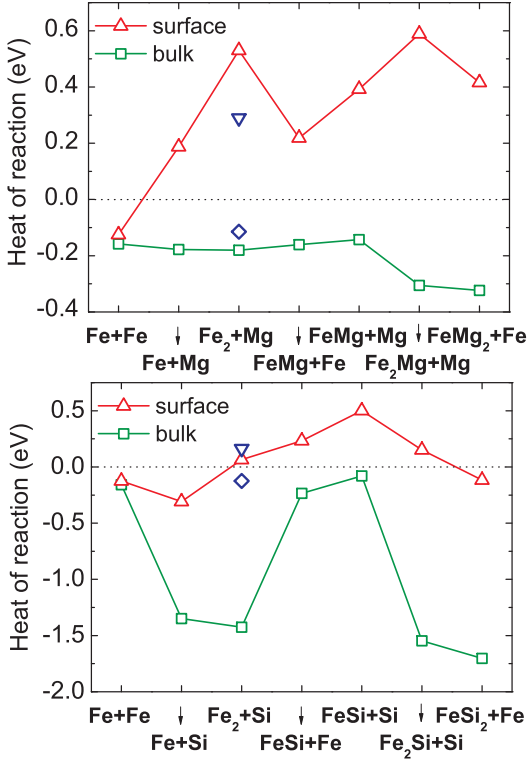}
      \caption{(Color online) Heat of reaction (pairing energy $E_{\text{d}}$) for the formation (at negative $E_{\text{d}}$) of successive Fe$_n$Mg$_m$  (upper panel) and Fe$_n$Si$_m$ (lower panel) nearest neighbor cation clusters ($n, m \leq 2$) in bulk zb-GaN (squares) and on the Ga-terminated (0001) wz-GaN surface (triangles). The solid lines are guide to the eye. The data points not connected by the solid lines (reversed triangles for surfaces, and diamonds for bulk) denote the heat of reaction of two nearest neighbor Fe cations in the presence of Mg or Si at the distance of 0.6 nm.}
     \label{fig:theory1}
\end{figure}

The formation of Fe pairs may be, however, strongly affected by the presence of Mg or Si atoms. This is shown in Fig.\,\ref{fig:theory1}, where we compare the heat of reactions for the formation of several consecutive Fe$_n$X$_m$ ($n, m \leq 2$) clusters, where X denotes Mg and Si, respectively.

\begin{figure}[ht]
		\centering
        \includegraphics[width=\columnwidth]{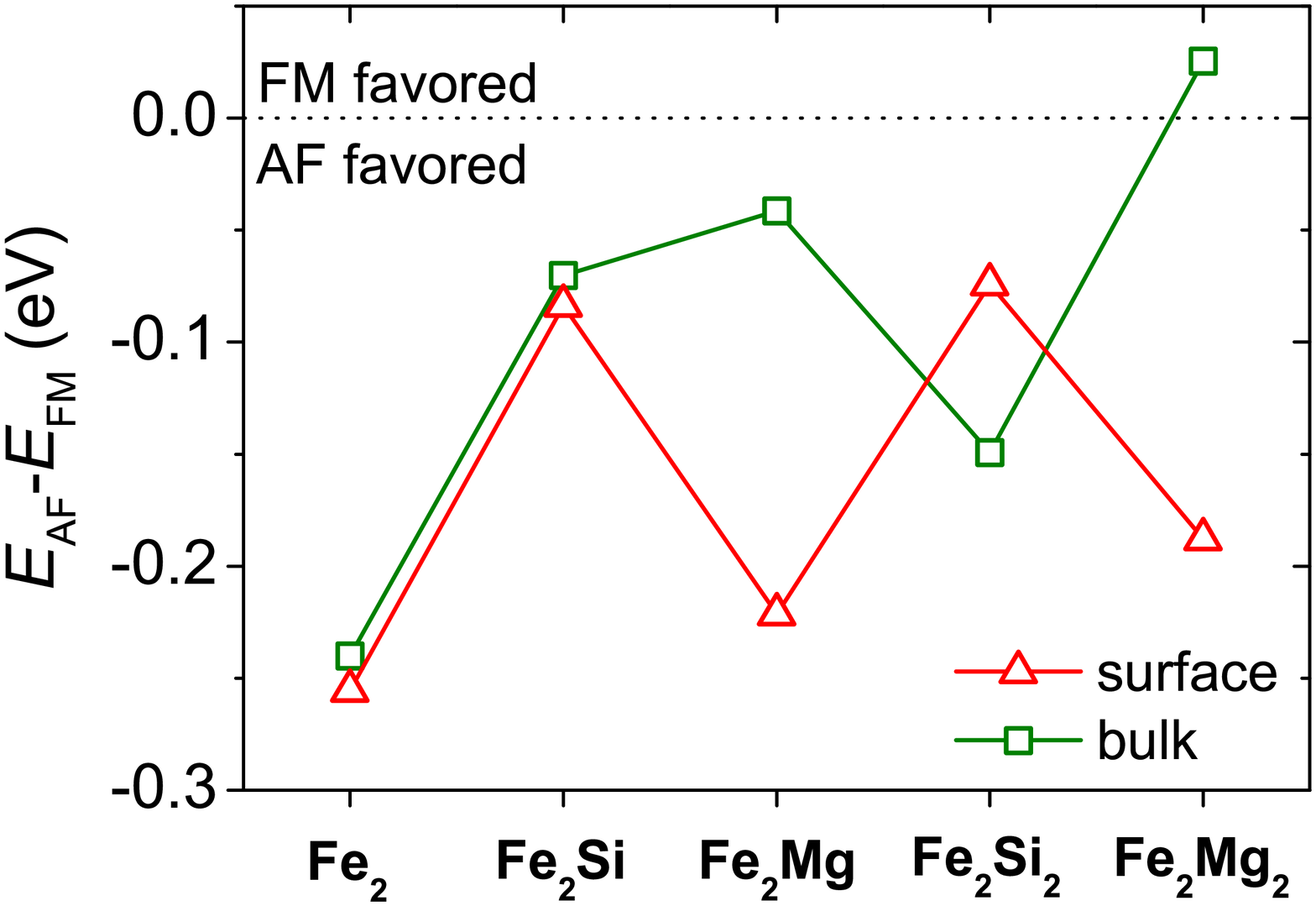}
      \caption{(Color online) Energy difference between ferromagnetic and antiferromagnetic configurations for several Fe$_{2}$X$_m$ ($m \leq 2$, and X denotes Mg or Si) clusters in zb-GaN and at the Ga-terminated (0001) wz-GaN surface. The solid lines are guides to the eye.}
     \label{fig:theory2}
\end{figure}

Codoping of (Ga,Fe)N with Si or Mg not only affects the formation of the Fe pairs, but also influences the magnetic coupling between the Fe ions, as shown in Fig.\,\ref{fig:theory2} where we plot the total energy difference between the ferromagnetic (FM) and antiferromagnetic (AF) configurations for several Fe$_2$X$_m$ (m $\leq$ 2, and X denotes Si or Mg) clusters in the zb-GaN matrix and at the (0001) wz GaN surface.

To complement our investigation, we have also carried out calculations for the formation of Fe$_n$Mg ($n \leq 2$) clusters in the presence of negative charge in the GaN matrix. The negative charge in the system may be viewed as an analog of the compensating defects ($e.g.$, H impurities) that can be present during the codoping process of (Ga,Fe)N with Mg. The heat of reaction Fe\,+\,Mg$^{-}$\,$\rightarrow$\,FeMg$^{-}$ is positive both in the bulk and at the surface ($E_{\text{d}}$\,=\,0.012 and 0.196\,eV for bulk and surface, respectively). The heat of the reactions Fe$_{2}$\,+\,Mg$^{-}$\,$\rightarrow$\,Fe$_{2}$Mg$^{-}$ and  Fe\,+\,FeMg$^{-}$\,$\rightarrow$\,Fe$_{2}$Mg$^{-}$, is 0.054 and -0.126\,eV, respectively, in the bulk, and 0.514 and 0.194\,eV, at the surface. This means that there is no qualitative difference between the results with and without charge at the surface. However, in the bulk a negatively charged Mg is not likely to be attached to Fe or Fe$_{2}$. For the hypothetical Fe$_{2}$Mg$^{-}$ the tendency to AF coupling between the Fe ions is preserved both in the bulk and at the surface.

Finally, we have also studied the possibility of the formation of Fe pairs at the Ga surface in the presence of Si or Mg in the proximity of a Fe ion, in our case at a distance of 0.6\,nm. The Fe ions -- that have a negative pairing energy at the clean surface -- avoid pairing in the presence of Si or Mg, since $E_{\text{d}}$\,=\,0.155 and 0.267\,eV, respectively. The AF coupling is still dominant for the hypothetical Fe pair in the presence of Si or Mg at the surface.

\section{Discussion and conclusions}

In this Section we discuss the outcome of the experimental studies summarized in Sec.\,III in the light of the theoretical results presented in Sec.\,IV. 

It is well known that except for isoelectronic Mn in II-VI compounds, where $d$-levels are far from the Fermi energy, in virtually all other transition metal (TM)-doped semiconductors a significant  contribution of open $d$ shells to the bonding leads to a strong attractive force between magnetic cations.\cite{Kuroda:2007_NM,Sato:2010_RMP} This leads to segregation and/or precipitation of TM-rich compounds during the growth under thermal equilibrium conditions, the resulting solubility limit being typically well below 0.1\%.  However, epitaxy at sufficiently low temperatures is known to allow for the fabrication of dilute magnetic semiconductors (DMSs) with random distribution of TM cations of concentrations largely surpassing the solubility limit, a canonical example being (Ga,Mn)As.\cite{Ohno:1998_S} The high temperature annealing of this material leads to the aggregation of Mn cations resulting in the formation of Mn-rich embedded ferromagnetic nanocrystals.\cite{De_Boeck:1996_APL} The ferromagnetic/semiconductor nanocomposite formed in this way shows the typical properties of a superparamagnetic systems.\cite{Sadowski:2011_arXiv} 

Extensive investigations over the recent years have made it increasingly clear that in most TM-doped semiconductors and oxides the aggregation of magnetic cations occurs already during the growth process.\cite{Bonanni:2010_CSR} The studies carried out to-date on TM-doped nitrides allow us to shed light onto a number of questions pertinent to these systems.

\subsection{Does the aggregation of magnetic ions occur in the bulk or at the surface?}

A strong dependence of the efficiency of Fe aggregation on the growth rate led us to conclude that this process takes place at the growing surface.\cite{Bonanni:2008_PRL} This fact was further supported by the lack of evidences for aggregation of both Fe and Mn under post-growth annealing up to above growth temperature.\cite{Navarro:2010_PRB,Stefanowicz:2010_PRB} At the same time, we found that under similar growth conditions, the solubility limit of Mn in GaN is at least one order of magnitude greater than the one of Fe.\cite{Bonanni:2010_arXiv} We explained this surprising difference within first principles computations\cite{Gonzalez:2011_PRB} implying that there is no attractive force between Mn cation pairs at the GaN surface, whereas the magnitude of the pairing energy of Fe cations $E_{\text{d}} = -0.125$\,eV,\ reported in Fig.\,\ref{fig:theory1} can explain the formation of the Fe-rich nanocrystals, even at a growth temperature of $\sim$1100\,K.

\subsection{Why does Mg, in contrast to Si, reduce the efficiency of Fe incorporation into the host?}

According to computational results displayed in Fig.\,\ref{fig:theory1}, there is a gain in energy with the formation of Fe-Si cation pairs at the GaN surface. In contrast, there is a repulsive force between Fe and Mg under the same conditions. This means, in line with the experimental results collected in Fig.\,\ref{fig:para} at low Fe flow rate, that the presence of Si enforces the Fe incorporation whereas Mg has an opposite effect, as observed. 

At the same time, the destructive effect of homogeneous Si and Mg codoping upon the Fe aggregation found experimentally in the previous\cite{Bonanni:2008_PRL} and present work -- as evidenced in Fig.\,\ref{fig:ferro} -- is in accord with the theoretically established repulsive interaction between Fe cations at the GaN surface in the presence of either Mg or Si, as shown in Fig.\,\ref{fig:theory1}. As discussed in Sec.\,IVB, the effect is stable against a donor compensation ($e.g.$, of Mg by H) and is independent of the distance of Mg or Si to the Fe-Fe pair. Once more it must be underlined, that in contrast to Si -- shown to rather promote the incorporation of Fe into GaN -- the homogeneous codoping with Mg diminishes the efficiency of the Fe incorporation into the GaN matrix, with the obvious effect of promoting the dilution of the system. 

It is worth noting that according to the results collected in Fig.\,\ref{fig:SIMS_Mg}, the presence of Fe favors the incorporation of Mg into GaN. This may mean that the strength of the repulsive interaction within Mg-Mg pairs is lowered by the presence of Fe. The {\em ab initio} results reported in Sec.\,IVB suggest that the pairing energies Mg-Mg and Mg-Fe are of the same order at the growth surface, 0.13 and 0.2\,eV, respectively. However, the presence of Fe leads to an attractive interaction within Mg-Mg and Mg-Fe pairs, that may promote the Mg incorporation, as observed.

\subsection{How does digital doping promote the formation of nanocrystals?}

For the case of digitally codoped layers: (i) in contrast to the case of homogeneous doping, the $\delta$-fashion growth-mode promotes the aggregation of Fe and the consequent phase-separation of the system; (ii) a high density of Fe-rich nanocrystals is observed already for 0.1\% of Fe under our growth conditions; (iii) the variety of the Fe-rich phases becomes more variegated with increasing Fe concentration; (iv) the nanocrystals are not distributed uniformly in the sample volume, but concentrate 300\,nm below the sample surface and at the interface GaN-buffer/(Ga,Fe)N:$\delta$Mg-layer; and (v) due to the presence of the Fe-rich embedded nanocrystals, a significant FM contribution to the overall magnetization is observed up to above room temperature. At the same time, the presence of Fe reduces the mean concentration of Mg that can be incorporated in the digital mode.

The origin of the above striking phenomena is not yet fully understood. It is possible that interdiffusion of Mg and Fe accompanying the digital growth mode makes that bulk rather than surface energetics applies to this case. According to Fig.\,\ref{fig:theory1}, the presence of Mg not only does not hamper the Fe aggregation but enhances it, particularly in the case of larger clusters. At the same time a substantial repulsion between Fe-Mg pairs at the surface may lead to a reduction of a mean Mg concentration in (Ga,Fe)N, compared to digitally-Mg doped GaN.

The overall effect of digital-doping and or/generally digital-growth requires at this point a dedicated study of growth-kinematics and -dynamics both in the presence and absence of intentional doping.

\subsection{Why are the embedded Fe-rich nanocrystals not distributed uniformly over the film volume?}

According to the present and previous\cite{Bonanni:2008_PRL} XRD and TEM studies, the average size of the embedded Fe-rich nanocrystals is independent of the Fe flow rate, suggesting that the precipitation occurs by a nucleation mechanism in which only nanocrystals with a critical size can form. If this is the case, during the epitaxy, as long as the nanocrystals maintain an undercritical size, they develop and segregate with the growth front, a process that is interrupted once the critical size is reached. Accordingly, a high density of nanocrystals is expected to be found in narrow volumes perpendicular to the growth direction. Within this scenario, the distance between the nanocrystals-rich volumes would be inversely proportional to the rate of Fe incorporation at the surface. The previous and present TEM data reveal the existence of such 2D-like distribution of nanocrystals. An important consequence of the resulting highly non-uniform distribution of the magnetic nanostructures over the film volume, is the onset of significant dipole-dipole interactions freezing the thermal fluctuations of the nanocrystals magnetic moments.\\
\\

\section{Summary and outlook}

The effect of Mg-doping on the structural arrangement and in the magnetic behavior of (Ga,Fe)N has been investigated, with particular attention to the role of the doping method (homogeneous or digital). The onset of a competitive mechanism  between Mg and Fe for the occupation of substitutional sites becomes evident, as the presence of Mg reduces the amount of Fe in the homogeneously Mg-doped layers compared to the corresponding reference samples (Ga,Fe)N and induces Fe aggregation in the digitally fabricated ones. As a consequence of the reduced efficiency of Fe incorporation in the homogeneously Mg-doped layers, the system becomes dilute and its magnetic response is purely paramagnetic. In contrast, upon $\delta$Mg-doping, the tendency of the Fe ions to aggregate is enhanced, and the system becomes ferromagnetic, due to the presence of FM $\varepsilon$-Fe$_3$N and \textit{galfenol} Fe$_{1-x}$Ga$_x$ nanocrystals.

The \textit{ab initio} calculations, confirm the tendency of the Fe ions to aggregate at the growing sample surface of wz-GaN efficiently in the presence of Mg. Furthermore, they indicate that: (i) the presence of either Mg hinders, while the one of Si promotes the Fe incorporation, as observed; (ii) both (homogeneous doping with) Mg or Si quenches the aggregation of Fe. This is in contrast with the computational results for bulk, where an enhancement of the tendency to aggregation is expected in the case of Mg codoping. It is possible that this case applies to the digital-like codoping, where interdiffusion is important.

In general terms, our combined experimental and theoretical studies demonstrate how particular codoping modes may serve to control the bottom-up fabrication of magnetic nanostructures in a semiconductor matrix. In view of our results, a considerable diversity of local atom arrangements obtained $via$ particular codoping modes results from substantial differences in the energetics at the growth surface and in the bulk, providing a worthy mean to design self-assembling processes at the nanoscale.

\begin{acknowledgments}
The work was supported by the European Research Council through the FunDMS Advanced Grant (\#227690) within the "Ideas" 7th Framework Programme of the EC, and by the Austrian Fonds zur {F\"{o}rderung} der wissenschaftlichen Forschung -- FWF (P18942, P20065 and N107-NAN). We acknowledge R.T. Lechner and G. Bauer for fruitful discussions and the technical staff at the Rossendorfer Beamline (BM20) and at the ``GILDA'' Beamline (BM08) of the ESRF, and in particular C. B\"{a}htz and N. Jeutter for their valuable assistance. The access to the computing facilities of the Interdisciplinary Center of Modeling at the University of Warsaw and of the High Performance Computing Center at Texas Southern University is also acknowledged.

\end{acknowledgments}

\bibliography{BibGaFeNMg}
\bibliographystyle{apsrev}

\end{document}